# Impact of gluconate and hexitol additives on the precipitation mechanism and kinetics of C-S-H

Lina Bouzouaid[1], Alexander E.S. Van Driessche[2], Wai Li Ling[3], Juan Carlos Martinez[4], Marc Malfois[4], Barbara Lothenbach[5], Christophe Labbez[1,*], Alejandro Fernandez-Martinez[2,*]

[1] *ICB, UMR 6303 CNRS, Univ. Bourgogne Franche-Comté, FR-21000 Dijon, France*
[2] *Univ. Grenoble Alpes, Univ. Savoie Mont Blanc, CNRS, IRD, IFSTTAR, ISTerre, 38000 Grenoble, France*
[3] *Univ. Grenoble Alpes, CEA, CNRS, IRIG, IBS, Grenoble, France.*
[4] *Non-Crystalline Diffraction Beamline−Experiments Division, ALBA Synchrotron Light Source, Cerdanyola del Vallès, Barcelona 08290, Spain*
[5] *Empa, Concrete & Asphalt Laboratory, Dubendorf, Switzerland*

\* Alex.Fernandez-Martinez@univ-grenoble-alpes.fr, christophe.labbez@u-bourgogne.fr

Abstract

The present paper investigates the influence of gluconate and hexitol additives on the precipitation mechanism and kinetics of C-S-H. To this end, wet chemistry C-S-H precipitation experiments were performed under controlled conditions of solution supersaturation –under varying silicate concentration, while the transmittance of the solution was followed. This allowed determining induction times for the formation of C-S-H precursors in the presence and absence of gluconate and three hexitol molecules. Characterization of the precipitates was performed via small angle X-ray scattering and cryo-transmission electron microscopy experiments, which allowed the identification of a multi-step nucleation pathway also in the presence of the organics. Analysis of the induction time data in the framework of the classical nucleation theory revealed a significant increase of the kinetic pre-factor, which is associated to physical rates of cluster collision and aggregation during the nucleation process. Values of the kinetic pre-factor increase in the same order as the complexation constants of calcium and silicate with the each of the organics. This points to a retarding mechanism of crystallization related to steric hindrance of the aggregation of the early formed clusters via adsorption of the organics at their surfaces.

1. Introduction

In Portland cement, calcium silicate hydrate (C-S-H) is the principal hydration product of alite ($C_3S$) and belite ($C_2S$), playing a decisive role in the final properties of concrete (1). It is generally accepted that the formation of this hydrate follows a dissolution-precipitation process (2) (3). To control the workability and final properties of concrete, a large variety of additives are added to the cement mixture, such as setting retarders, accelerators and superplastizicers, among others (4) (5) (6) (7). These chemicals alter different processes occurring in cementitious systems, including their hydration behavior. Extensive experimental work has shown that numerous additives have a retardation effect on the hydration of $C_3S$ (8) (9) (10) (11) (12) (13), mainly by disrupting the nucleation and/or growth process of C-S-H. In addition, it has been shown that the dissolution kinetics of cement and/or $C_3S$ are not limited by different retarders, such as yellow dextrin and cellulose ethers (12) (14) (15) (16) (17). Thus, in order to develop more efficient cement additives, a detailed understanding of the initial formation of C-S-H in the absence and presence of these additives is needed.

In a recent work by Krautwurst et al. (18), time-resolved potentiometry coupled with transmittance measurements, small angle X-ray scattering and cryo-TEM experiments were used to study the homogeneous nucleation of C-S-H from diluted solutions. It was found that C-S-H formation follows a two-step pathway. The first step, identified by an decrease of the transmittance of the solution, corresponds to the formation of amorphous spheroids which subsequently crystallized into typical sheet-shaped C-S-H particles during a second step of transmittance decrease (18). In another study, Plank et. al. used TEM imaging to show the initial presence of metastable C-S-H nanoparticles with globular morphology and sizes between 20 and 60 nm, similar to those observed by Krautwurst et al. (18). After less than one hour, those nanoparticles were observed to grow gradually into the characteristic nanofoils of early C-S-H.

These two studies described the formation of C-S-H through a so-called 'non-classical' nucleation pathway, which has been observed for a large variety of organic and inorganic systems, but remain a hot topic of debate and investigation due to the complex processes involved (19) (20) The formation of precursor metastable phases is predicted by Ostwald 's rule of stages, invoking the low energetic cost from a kinetic point of view to form the less stable polymorph (i.e. more soluble) from the supersaturated solution. Many recent observations have contributed lately to a more complete description of this rule of stages: several authors have described the formation of amorphous or nanoparticle precursors (e.g. (21) (22)), which in some cases act as the building blocks of the crystalline phases through complex diffusion and aggregation processes which are still poorly understood (e.g. (23) ). These observations have been validated for some systems, but are still far from being fully understood, and a universal law describing the full crystallization process is still far from being reached (e.g. (24)). Moreover, Plank et al. showed that polycarboxylate ether-based superplasticizers (PCEs) control the kinetics of the transition from globular to nanofoil-like C-S-H. Indeed, in the presence of those additives, this conversion is strongly delayed. It is argued that PCEs form layers around C-S-H globuli in a core-shell geometry. The authors proposed that this external layer slowed down the water diffusion into the particles' core, and consequently decreases the kinetics of dissolution and re-crystallization of the core particles. Yet, most of the commonly employed additives are small simple organics, such as gluconate, which is routinely used by the concrete industry as retarder to control the setting time of cement. (25) (26) (27) (28). Gluconate is also efficient in increasing the compressive strength to concrete (29) (30). Another saccharide derivative such as sorbitol (a neutral sugar alcohol) is also used as a set retarder for Portland cement: when added in the material at 0.40 wt%, sorbitol delays the setting of 2 days (31). Moreover, sorbitol is used as a water-reducing plasticizer, decreasing final porosity and giving greater mechanical strength and durability (32).

In overall, the case of C-S-H nucleation from diluted solutions is an archetypical case of multi-step nucleation, in which a metastable phase (amorphous spheroids) is persistent before its crystallization takes place. The aggregation of these spheroids has been suggested as a critical step, concomitant to the crystallization, but further investigations are needed to fully characterize the precipitation process, especially in the presence of additives. Therefore, the focus of this work is on the interactions of the amorphous C-S-H spheroids with different cement additives, D-sorbitol, D-mannitol, D-galactitol and of gluconate, that act as retardants of the crystallization process. The first three molecules share the same chemical formula but have different stereochemistry. Gluconate is a charged molecule known to act on the nucleation kinetics and mechanisms of C-S-H. We worked in diluted aqueous solutions and the pH of the experiments was set at the value close to the one observed in cement paste (pH 12.8). As opposed to the conditions reigning in real cement, and with the objective of simplifying the study, our goal was to study homogeneous nucleation, so the solutions were kept free from grain/impurities, in order to avoid any heterogeneous nucleation. First, the effects of additives on the nucleation kinetics of C-S-H were studied by determining the induction time as a function of supersaturation using transmittance measurements. Secondly, small angle X-ray scattering (SAXS) experiments were performed in order to characterize the size and shape distribution of the C-S-H nanoparticles formed during precipitation, in absence and in presence of additives. Finally, cryo-TEM, was used to get a clearer picture of the particle morphology in the presence and absence of organic molecules.

## 2. Material and methods

*2.1. Materials*

The different stock solutions were prepared by dissolving $CaCl_2$ $2H_2O$ (Sigma-Aldrich, ≥99% purity), sodium gluconate ($C_6H_{11}NaO_7$, Sigma-Aldrich, ≥99% purity) D-sorbitol ($C_6H_{14}O_6$, Sigma-Aldrich, ≥99% purity), D-mannitol ($C_6H_{14}O_6$, Sigma-Aldrich, ≥99% purity), and D-galactitol ($C_6H_{14}O_6$, Sigma-Aldrich, ≥99% purity) and sodium chloride (NaCl, Sigma-Aldrich, ≥99% purity), in boiled and degassed milliQ water.

The hexitols used in this study are all isomers, sharing the formula, $HOCH_2(CHOH)_2CH_2OH$, but differ in the stereochemical arrangement of the OH groups as illustrated in Figure 1 .

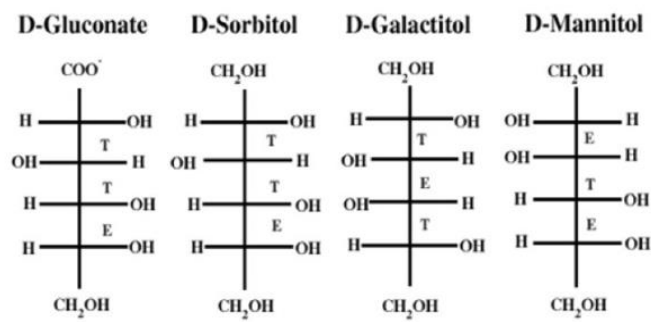

**Figure 1.** Stereochemistry of the different organic molecules used in this study.

*2.2. Titration of $Ca^{2+}$ with $Na_2SiO_3$,*

The precipitation of $CaO\text{-}SiO_2\text{-}H_2O$ was followed using an automated titrator instrument (Metrohm 905 Titrando) . All measurements were performed in a thermostated reactor at $23.0 \pm 0.1°C$. The solution in the reactor was continuously stirred with a magnetic stirrer at a constant rate of 430 rpm. A nitrogen flow circulated continuously above the solution to avoid contamination with $CO_2$. It was taken care that the gas did not enter the solution to avoid any disturbance of the electrodes. In the reactor, 148 mL of a solution containing 25mM NaCl background electrolyte, a fixed concentration of the organic of interest, between 0.1mM and 10mM, and 15 mM of $CaCl_2.(H_2O)_2$ After 3 minutes of equilibration time, 3ml of a titrant solution containing a silicate concentration is added. The final silicate concentrations reached were situated between 0.05mM and 0.25mM at a rate of 1.2ml/min registered in the software tiamo 2.5 (Metrohm). The addition of NaOH within the reactor allowed us to set the value of pH at 12.7-12.8, the final concentration of NaOH being fixed at 80mM.

In order to follow the formation of the first particles of C-S-H in absence and in presence of the small organics, an optical electrode (Metrohm Optrode, 6.1115.000) measuring the light transmittance of the system is added. A stable electrode signal is obtained with the use of a background electrolyte. The titrations were thus performed in 25 mM NaCl to ensure a stable signal and to limit the influence of the background electrolyte on the complex formation. One particular wavelength (among eight) using the software tiamo may be selected. For the experiment, the wavelength at 470 nm is chosen.

The calcium potential was also measured during the experiment, using the ISE $Ca^{2+}$ electrode (Metrohm Ca ISE 6.0508.110), coupled to a reference electrode (Metrohm Ag, AgCl/3 M KCl, 6.0750.100). By plotting the measured mV against the calculated $Ca^{2+}$ activity calculated with PHREEQC as detailed below, the calcium activity can be extracted. The pH was determined with a pH electrode (Metrohm pH Unitrode with Pt 1000, 6.0259.100), which allows reliable measurements up to pH = 14. The pH electrode was calibrated prior to the measurements with standard buffer solutions (pH 10, 12.45 and 14 from Sigma Aldrich).

*2.3. Thermodynamic simulation: saturation index calculation.*

The saturation index (SI) was calculated with a speciation model solved by the geochemical software PHREEQC version 3 (3.6.2-15100) using the WATEQ4f database. SI is calculated by comparing the activity product of the dissolved ions of C-S-H (IAP) with their solubility product ($K_{sp}$); SI = log(IAP/$K_{sp}$).

*2.4. Cryogenic Transmission Electron Microscopy*

Titration experiments, as described in 2.2, were run in the presence of gluconate and sorbitol and at different time points of the reaction 4μl aliquots were retrieved from the solution in the reactor and applied to freshly glow discharged qauntifoil or lacey carbon grid and vitrified using a Thermofisher Vitrobot Mark IV system. The grids were then mounted on a Gatan 626 single-tilt cryo-transfer holder. Imaging and selected area diffraction (SAD) were performed under low-dose conditions on a Tecnai F20 microscope operating at 200 keV. Images were recorded on a Ceta CMOS camera whereas diffraction patterns were recorded on an Amsterdam Scientific Instrument CheeTah hybrid pixel camera.

*2.4. Small Angle X-ray Scattering (SAXS)*

SAXS experiments were performed at the NCD beamline of the ALBA synchrotron (Barcelona, Spain). The same Metrohm system described in section 2.2 was used to titrate a calcium chloride solution containing gluconate with a sodium orthosilicate solution (see concentrations used in Table 1). The reaction took place in the same reactor used for the transmittance experiments described in section 2.2. A peristaltic pump was used to circulate the solution in a closed loop through a 1.5 mm diameter polyimide capillary that was set up at the beam position. The residence time in the tubing and capillary is in the order of ~20 seconds. The chemistry was followed using a pH electrode and a transmittance sensor (as described in 2.2). An X-ray beam of energy E = 12.40 keV ($\lambda$ = 1.00 Å), calibrated using a silver behenate standard, was used. The scattered radiation was collected using a Pilatus 2M two-dimensional detector (981 × 1043 pixels, 172 × 172 μm pixel size) that was placed at 6.7 m from the sample to yield a q-range 0.03 < q < 2.00 nm$^{-1}$. The scattered intensity was acquired every minute in acquisitions of 30 seconds to follow the kinetics of C-S-H nucleation.

**Table 1.** Conditions used for the SAXS experiments.

| Additive | [Si] (mM) | [Ca] (mM) | [NaCl] (mM) | [NaOH] (mM) | [additive] (mM) |
|---|---|---|---|---|---|
| Pure | 0.10 | 15 | 25 | 18 | - |
| Gluconate | 0.10 | 15 | 25 | 18 | 0.03 |
| Gluconate | 0.10 | 15 | 25 | 18 | 0.07 |
| Gluconate | 0.10 | 15 | 25 | 18 | 0.15 |
| Gluconate | 0.10 | 15 | 25 | 18 | 0.30 |

## 3. Results and discussion

*Thermodynamics of C-S-H nucleation*

Three curves showing the typical evolution of the transmittance over time, for the pure and the 10 mM gluconate and sorbitol systems, are shown in Figure 2. The pure system shows the same behaviour described by Krautwurst et al. (18), with two characteristic drops of the transmittance. The first drop was described as corresponding to the formation of amorphous spherical particles of ~50 nm size that act as precursors for the formation of nanocrystalline C-S-H; the second drop was described as resulting from a process of droplet aggregation and changes of stoichiometry and surface chemistry, leading to massive C-S-H crystallization (18). This behaviour is also observed in the case of sorbitol, showing two well-defined drops of the transmittance. Importantly, the case of gluconate is notably different: the signal decrease is slower than for the pure system and for the sorbitol and no clear sign of a second drop of the signal at later times is observed.

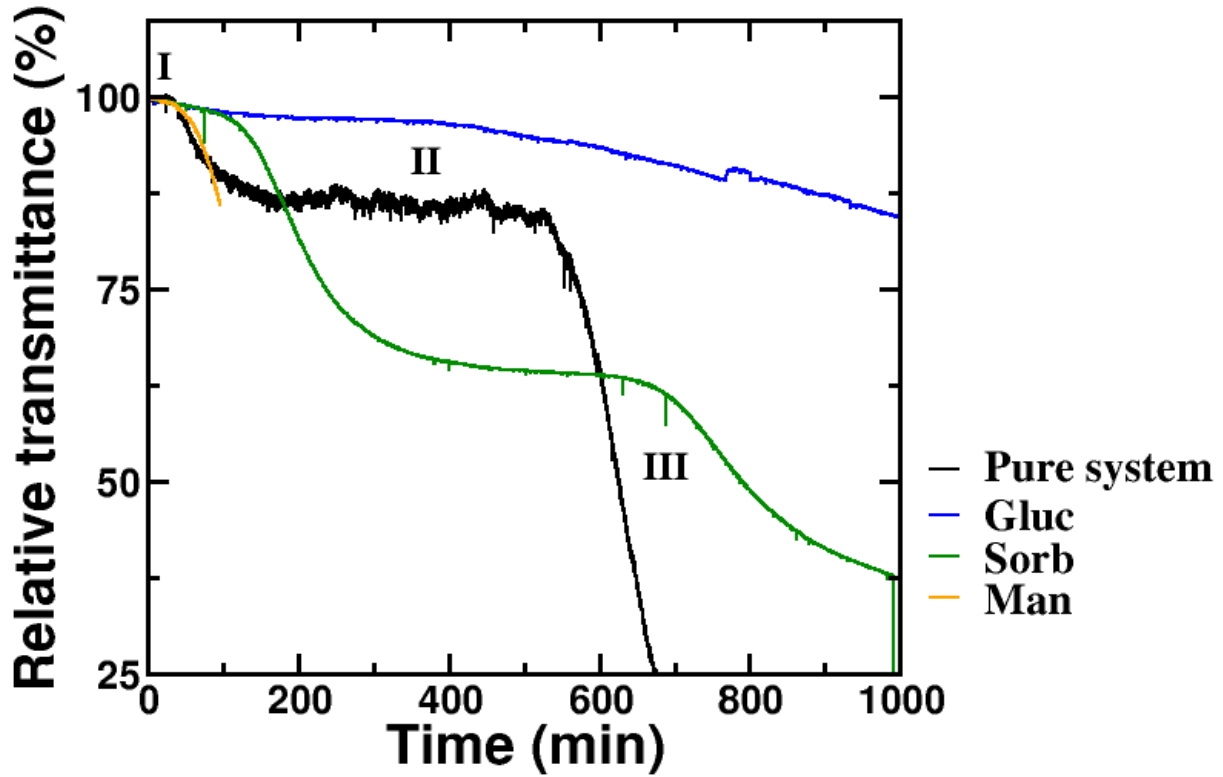

**Figure 2**. Evolution of the relative transmittance of the solution as a function of time for pure system of precipitating C-S-H as reference experiment (black line), and in the presence of 10 mM gluconate (blue), 10 mM sorbitol (green), 10 mM mannitol (orange), and 10 mM galactitol (pink). The final concentrations of Si is set at 0.10 mM, Ca at 15 mM. The total solution volume is set at 150 ml. After an initial drop of the transmittance (stage I) corresponding to the formation of amorphous droplets, a plateau is reached (stage II). A final decrease of the transmittance is observed (stage III), which has been assigned to C-S-H crystallization.

A detailed analysis of the initial times shows a difference between the behaviour of the pure system and that of the organic-containing systems. The neutral molecule sorbitol and the charged molecule gluconate are represented as examples (Figure 3, center and right), and compared with the reference system (Figure 3, left; data are for solutions with the same organic concentration of 10 mM and same $Si_{final}$ = 0.15 mM and $Ca_{final}$ = 15mM). Whereas the data from the pure system shows an initial plateau, with no changes in the transmittance until the induction time, the solutions containing organics show an initial decrease of the transmittance that takes place immediately after mixing the solutions, followed by a stabilization step prior to a second drop. This initial decrease is more pronounced in the case of gluconate than for sorbitol: in the case of sorbitol, it lasts for less than 7 minutes, and the drop in relative transmittance is less than 1%; the data from the gluconate-containing solution shows an initial decrease that lasts for ~100 minutes solution, and a drop in relative transmittance of ~2%.

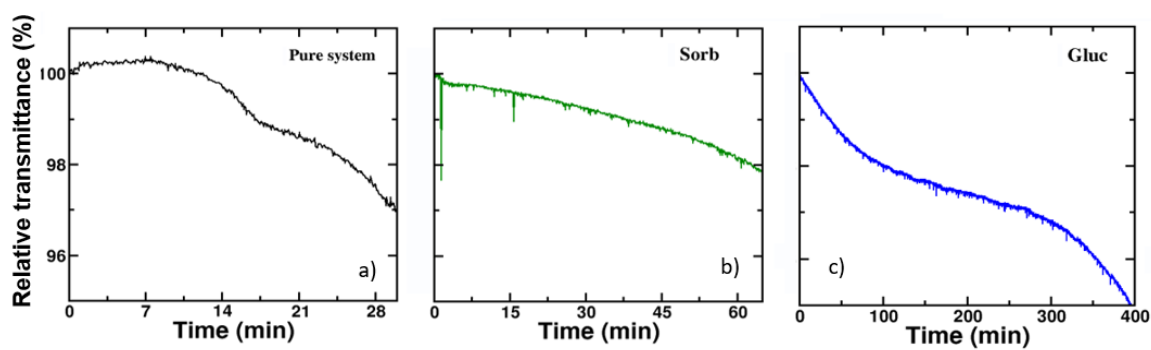

**Figure 3**. Enlargement of the initial part of the transmittance curves, before the first transmittance drop of stage I, with $Si_{final}$=0.15mM and $Ca_{final}$=15mM. 3.a: Transmittance as a function of the time for the pure system. 3.b: Addition of sorbitol at 10mM in C-S-H system. 3.c: Addition of gluconate at 10 mM in C-S-H system.

The presence of this first drop in the organic-containing solutions poses some questions about the interpretation of the different phenomena going on during the C-S-H nucleation in the presence of additives. In the analysis by Krautwurst et al. (18) this very initial first drop in transmittance was not observed as confirmed here, see Figure 3-a. The fact that this initial drop is not observed for the pure system allows making the hypothesis that it is related to a process of formation of aqueous Ca-Si-organic complexes. With the aim of clarifying this point, a comparison of data taken for the 10 mM gluconate system under different β-C-S-H supersaturations is presented in Figure 4 (see speciation of the supersaturated solutions in Table S1.A of the Supporting Information). The data show (Fig. 4a) that all the solutions display the same trend, with an initial drop of the transmittance in the same time interval, irrespective of the supersaturation with respect to β-C-S-H. Such a behaviour differs notably from that expected for a nucleation event, in which the induction time –and its associated decrease of the transmittance– responds to changes in the supersaturation of the solution. Indeed, the initial drop is independent of the supersaturation. The associated process is thus not activated. On the contrary, the second drop in transmittance(corresponding to the first drop in the pure system), observed at around ~500 min shows a dependency with the supersaturation, with faster and more pronounced decrease at higher supersaturations. This leads us to propose that the observed the very first drop is due to the above-mentioned process of aqueous Ca-Si-organic complex formation (or complex clusters) and that the second drop at >500 min (Fig. 4b) is a true nucleation event. Note that the first two values (Si = 0.08 mM and Si = 0.10 mM) do not follow the expected trend; this behaviour, though unexpected, can be rationalized considering the stochastic nature of the nucleation process, which leads to non-negligible uncertainties in the values of the induction times, as it will be seen later.

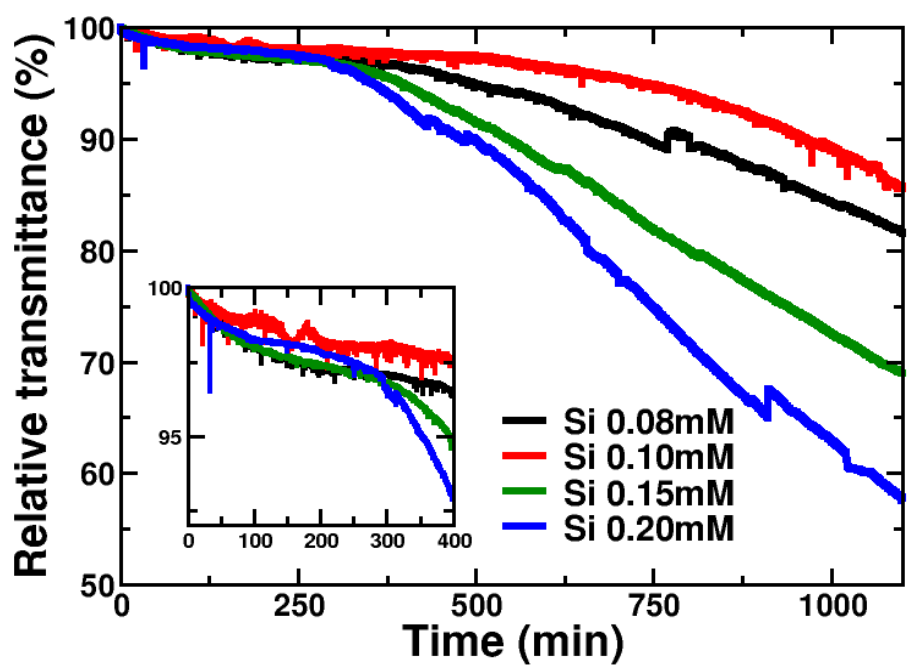

**Figure 4**. Transmittance as a function of time for the C-S-H system in presence of gluconate at 10mM, at different supersaturation degrees: [Si]=0.08mM, [Si]=0.10mM, [Si]=0.15mM and [Si]=0.20mM.

The data shown in Fig. 4 allows assigning the initial drop in transmittance to a process of Ca-Si-organic complex formation. It seems that the organic molecules interfere with the nuclei precipitated, or seem to induce the formation of particular nuclei that are different in shape and size, blurring the well-defined stages observed in the reference system, in particular stage I. The presence of the complexes involving calcium and the organics described in more details in our previous work (33) can be an explanation in the change of the pathways observed for the pure system, and thus can be the cause of the very first decrease of the transmittance, given that it is observed at all supersaturation degree studied (see Figure 4). The fact that, as shown in Fig. 3, this initial drop is larger for gluconate than for sorbitol is expected given the relative amount of calcium being part of multinuclear Ca-Si-gluconate complexes that form under these conditions, in comparison to the lower amount in hexitol complexes (see Table S1.B of the Supporting Information). (33).

Thus, the later drop (at ~500 min in Figure 4) can be considered as the first nucleation event occurring in the system and allows, us to perform a quantitative analysis of the thermodynamics and kinetics of nucleation using classical nucleation theory (CNT) [e.g. (34)]. To this end, all the transmittance curves were analysed by doing linear fits as explained in Figure S1 (see Supporting Information). Following the CNT formalism, the induction time ($t_{ind}$) can be written as a function of the solution saturation index ($SI$):

$$\log(t_{ind}) = C_0 + \frac{\frac{16}{3}\pi \Omega^2 \alpha^3}{(k_B T)^2 SI^2} \qquad \text{(eq. 1)}$$

where $\Omega$ is the molar volume of the formed phase, $\alpha$ is the interfacial free energy, $k_B$ is the Boltzmann factor, and T is the temperature. $C_0$ is the so-called pre-exponential kinetic factor, and it is related to thermodynamic and kinetic factors such as the number of sites susceptible to act as nucleation loci, the collision rate, and the Zeldovich factor (related to the probability of successful formation of a cluster in solution). Induction times from all the experiments performed are shown in Figure 5.

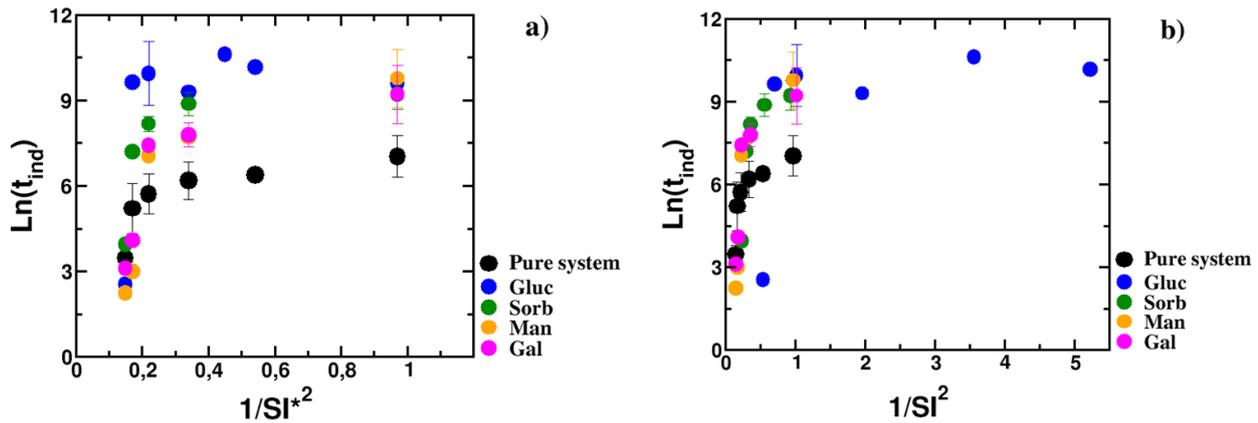

**Figure 5.** Logarithm of the induction time as a function of the saturation index for the pure system (black) and in presence of gluconate (blue), sorbitol (green), mannitol (orange) and galactitol (pink). The saturation index was calculated using the speciation software PHREEQC, taking into account the condition of the different precipitation experiments, including components concentrations and temperature, but also the complexes existing between calcium, silicate and the organic molecules (b). The apparent SI* was also calculated, which does not take into account the aqueous complexes involving the organics (a).

As expected, the overall trend for all the datasets is an increase of the induction time when the saturation index decreases. An overall increase of the induction times when organic molecules are added to the solution can also be observed. This increase is pronounced in the range of lower supersaturations (high $1/SI^2$ values). Data at higher supersaturations blur this effect as the induction times decrease, with the statistical error bars being more significant.

To fit these data using eq. 1 some approximations need to be made: the first one is the assumption that the nucleated phase is β-C-S-H. Indeed, the supersaturation values used to plot the data have been calculated using the solubility product with respect to that phase, and the only value for a molar volume available in the literature is that corresponding to a nanocrystalline C-S-H phase such as β-C-S-H. However, as discussed above and reported by Krautwurst et al, and as it is confirmed by our cryo-TEM data, the drop in transmittance used to determine the induction time corresponds to the formation of amorphous droplets. In the absence of precise information about the density and solubility of these droplets, the approximation is made that β-C-S-H is the nucleated phase. This implies that the values for the interfacial energies used cannot be interpreted in absolute terms, but only relatively to each other. A second approximation made, inherent to the use of the master equation of CNT in eq. 1, is the fact that the same nature of precipitate is formed for all the supersaturation conditions. However, data fitted using all the supersaturation range and a supersaturation dependence of the pre-exponential term ($C_0$) shows a deviation at high supersaturation values, which probably points to a change in the nature of the precipitate at high Ca and Si concentrations (see Figure S1). A way to fix this issue is to separate the experimental data in two parts: low and high supersaturations. Here, data from low supersaturation has been fitted using eq. 1, and considering $C_0$ as a constant independent on the value of the saturation index, an approximation widely used in the literature.

Results of the fits yield values for the interfacial free energy or surface tension, $\alpha$, and of the kinetic pre-factor, $C_0$ (Fig. 6). All the values of the interfacial free energies, $\alpha$, fall in the range between 5 and 13 mJ/m$^2$, In agreement with the value (~13 mJ/m$^2$) obtained by Gauffinet and Nonat (35) with large uncertainties that make them indistinguishable within the error. The situation changes for the values of the kinetic pre-factor, $C_0$. The gluconate system shows the higher values of $C_0$. The systems with the hexitols sorbitol, galactitol and mannitol yield a similar value of $C_0$, which is still slightly higher than that of the pure system. Interestingly, this order matches that found for the complexation constants of $Ca^{2+}$ and silicates with these organics: gluconate was found to form strong polynuclear complexes with $Ca^{2+}$ and silicates, followed by sorbitol, which showed a weaker complexation level, and than by galactitol and mannitol, with the weakest tendency to form complexes.

The fact that the presence of the organics does not modify (within the uncertainty) the value of the interfacial free energy would suggest that the nature of the C-S-H nuclei formed –their interfacial properties remains rather the same irrespective of the presence of the organic molecules. In that case, a plausible interpretation for all the data in Fig. 6 would consist in a process by which the organics interact physically with the nucleated particles, modifying their aggregation pathways. Indeed, as suggested by Krautwurst et al. from observations with the transmittance probe, the individual, precursor particles undergo crystallization at stage III, in a crystallization process that is concomitant to the aggregation of the amorphous spheroids. This type of effect, where an organic additive imposes steric barriers to the aggregation, modifying the crystallization kinetics, has already been observed by other systems such as $CaSO_4$. (36) However, the interfacial free energy data have to be analysed carefully. The reason for this is that the nucleation rate is highly sensitive to the value of the interfacial free energy due to the cubic exponent in eq. 1. This fact makes that a small variation could change the height of the nucleation barrier (and therefore the nucleation kinetics) in a very significant way. This is exemplified in Figure 7, where a sensitivity analysis of this parameter, the interfacial free energy, is shown. Values of the height of the nucleation barrier –normalized by the thermal energy, $k_BT$- are given in the colour bar, and are plotted as a function of the interfacial free energy ($x$ axis) and the saturation index of the solution ($y$ axis). The range of values in both $x$ and $y$ axes have been set to values that are used or found in the experiments reported here. The results show that the nucleation barrier can vary between very small values (< 5) where nucleation can happen spontaneously, to values in the order of ~30 at which induction times for nucleation tend to infinite. This sensitivity analysis leads us thus to make a cautionary note about the results of the interfacial free energies: the relatively large error bars precludes any detailed discussion about the implications of the values found in terms of the height of the nucleation barrier.

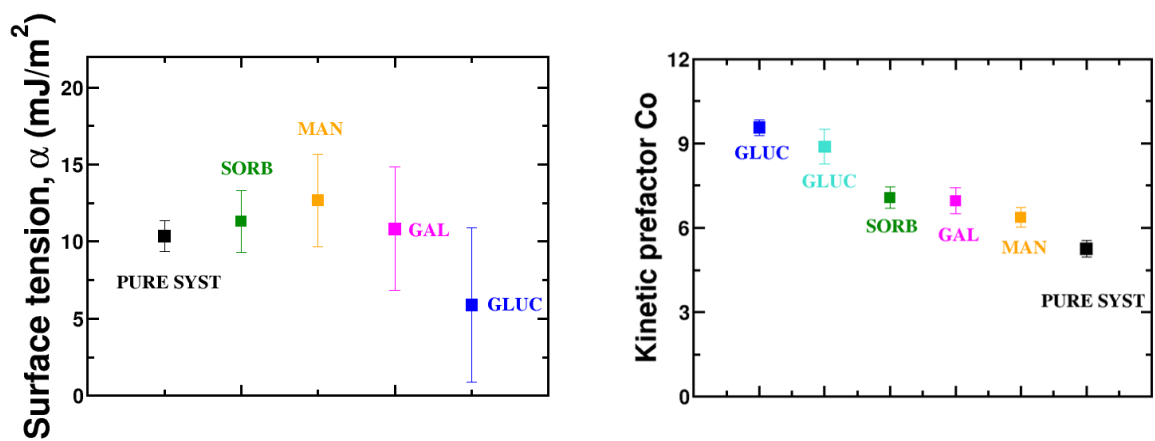

**Figure 6.** Left: Values of the interfacial free energies, α, obtained for the pure and organic-containing C-S-H systems. Right: Values of the kinetic pre-factor, $C_0$. The concentration of the organics is 10 mM. The light blue color used for gluconate in the right figure, gives $C_0$ as obtained for a concentration of 1 mM.

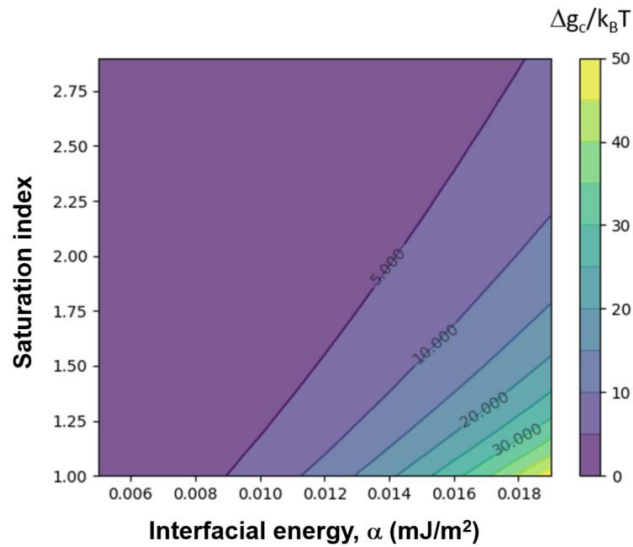

**Figure 7.** Height of the nucleation barrier (in $k_BT$ units) as a function of the interfacial energy and saturation index of the solution with respect to β-C-S-H.

*Physical characterization of the precipitates*

Cryo-TEM images of the precipitates formed in the presence of 1mM gluconate and 5mM sorbitol, each imaged at different times during the nucleation process, are shown in Figure 8. Particles show a spherical morphology in all cases, with larger particles (~500 nm – 1 μm) formed in the presence of gluconate. The size of the particles formed in the presence of sorbitol is smaller, ~10-50 nm, similar to the sizes found by Krautwurst et al. for the pure system under identical chemical conditions of solution supersaturation (18). The electron diffraction patterns shown as insets in Figure 8 reveal that the particles formed are amorphous.

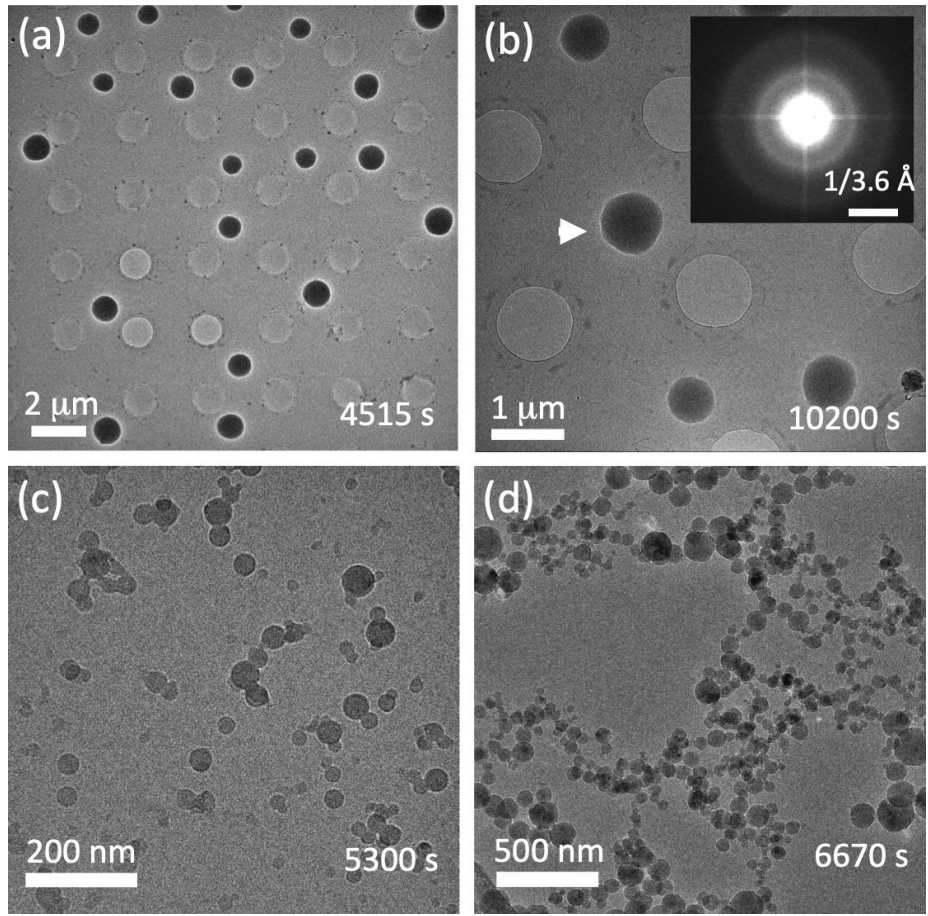

**Figure 8.** Cryogenic TEM images of the early stages of C-S-H precipitation in the presence of 1mM gluconate (a,b) and 5 mM Sorbitol (c,d). White arrow in (b) indicates the particle that was used for selected area diffraction (SAED) shown in the inset of (b). Times elapsed between the onset of the reaction and the cryo-quenching of an aliquot of the reacting solution.

SAXS data for some selected times are represented in Figure 9. In all cases, an increase of the intensity is observed over time. However, the pure system exhibits a clear induction time, with the intensity increasing only after 10 mins, whereas no induction time is observed for the gluconate-containing systems (see evolution of the SAXS invariant in Figure 9). This is in agreement with the turbidity data shown before, where the formation of the complexes with gluconate induced an immediate decrease of the transmittance upon mixing of the silicate and gluconate-bearing calcium solutions.

SAXS patterns from the pure system evolve from a particle scattering characteristic of a very polydisperse system (30 mins), with no characteristic form factor, to a system where the intensity distribution follows a $q^{-2}$ dependence, as expected for plate-like systems. It is important to keep in mind that the SAXS is sensitive to density variations, and that these can also take place within the amorphous droplets, which are very hydrated systems. A change of slope is observed at ~$3 \cdot 10^{-1}$ nm$^{-1}$, which is tentatively ascribed to the thickness of the plates, which would be in the order of 20 nm. However, the polydispersity of the data precludes a detailed analysis of the morphology. It is interesting to observe that the plate-like particles are observed here after ~60 mins. According to Krautwurst et al., the particles are still amorphous at this time. We make here the hypothesis that plate-like atomic arrangements are adopted by silicate molecules and calcium atoms within the amorphous spheroids (identified as silicate dimers by Krautwurst et al.), and that these happen as a preliminary step of the crystallization. The low concentrations used in these experiments made that no signal, at any point, was observed in the wide-angle detector available at the SAXS beamline.

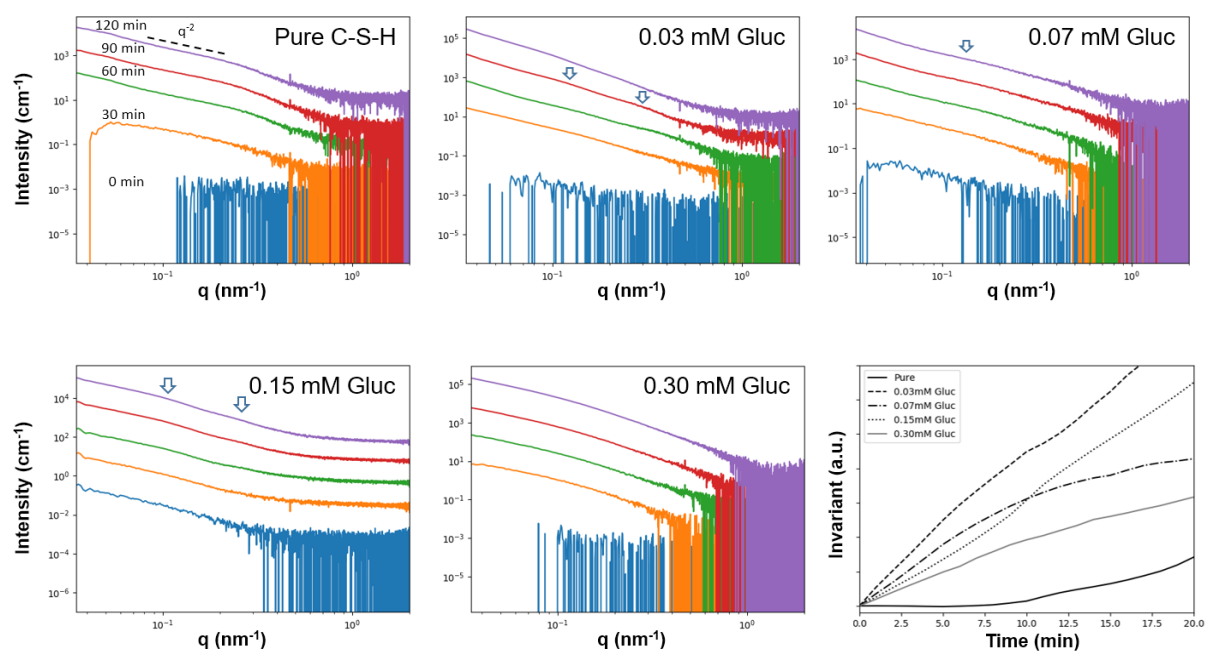

**Figure 9.** SAXS patterns and evolution of the SAXS invariant over time for the pure and gluconate-bearing systems. The arrows indicate bumps in the intensity coming from particle scattering within the precipitates.

Data from the systems containing gluconate are more complex. Different small bumps (see arrows in Figure 9) are observed that could indicate the presence of particles being aggregated into large networks. Indeed, the low $q$ part of the data follow a power law in almost all cases (except for the system with 0.30 mM gluconate), which can be fitted using a $I \propto q^{-n}$ law, as shown in Figure S2. The results show that the values of the exponent tend towards values higher than 2 in almost all cases , which are typically associated to mass fractals aggregates, tending to surface fractal when the values are higher than 3 (see Figure S2).. The case of the system with 0.30 mM of gluconate is different from the rest: all the data exhibit a certain degree of curvature, which again can be explained by a very polydisperse system with no particular form factor, as it was the case for the early stages (~30 mins) of nucleation of the pure system.

Another observation that is worth mentioning is the fact that the background signal is significantly higher in the system with 0.15 mM gluconate than in the rest. This could point to the presence of very small entities (sub-nm) that could be present in the solution (ion pairs, small complexes..). The nature of these cannot be ascertained from these data. In overall, the cryo-TEM and scattering data further confirm that organics do not alter the mechanism of the nucleation pathway, but rather slow down the different processes taking place en route from dissolved species to the final solid phase.

## Conclusions

The main conclusions that emerged from this study can be summarized as follows: the precipitation and crystallization step with the characteristic drops of transmittance, already observed by Krautwurst et.al. (18) were confirmed here for the pure C-S-H and they were also found for the systems containing organic molecules. Cryo-TEM observations confirmed that a nucleation pathway through an amorphous precursor takes place also in the presence of gluconate and sorbitol. The analysis of the induction times using classical nucleation theory yields similar values for the interfacial energies, within the uncertainties. This points to similar precursor phases (not chemically different), though the results have to be taken with caution given the large error bars and the sensitivity of the induction times to the value of the interfacial energy. The analysis of the pre-exponential factor, a kinetic factor that includes terms related to the physics of the collisions between clusters formed during the nucleation process, yields an interesting observation: the values of this factor increase from the pure system to the gluconate-containing systems, and are directly proportional to the values of the complexation constants found for each of the organics with the C-S-H phase in a previous study (the paper related to the matter will be submitted to Cement and Concrete Research soon) This observation may suggest that the mechanism by which the organics delay the nucleation is related to their ability to hinder the aggregation of clusters (preventing their aggregation); or that it is related to the organics stabilization of the pre-nucleation clusters of silicate and calcium, slowing down the formation of the metastable amorphous spheroids. This is in line with the report by Krautwurst et al. (18) by which i) the amorphous spheroids are formed and ii) the processes of particle aggregation and C-S-H crystallization are intimately related. SAXS analyses bring a physical perspective about the mesoscale organization of the particles: primary particles within the amorphous droplets are observed, adopting a plate-like configuration prior to their crystallization of pure C-S-H. This phenomenon is less clear for the organic-bearing systems, where a more complex shape of the curves suggest the formation of primary particles that aggregate forming larger clusters of sizes closer to those observed by cryo-TEM. Finally, cryo-TEM shows the formation of much bigger particles in presence of gluconate, compared to the amorphous droplets of pure C-S-H or of sorbitol-bearing C-S-H. This can be explained by the much higher complexation constant of gluconate (33) than for sorbitol, which would not be affecting the aggregation process at a large extent.

In spite of the interesting observations found here, it is unknown whether the amorphous spheroids are a requirement for C-S-H formation in real cement pastes. Experimental studies highlighted that C-S-H formation can be described as primary heterogeneous nucleation and growth (37) (38) in the real cement conditions. Indeed, given that dislocations, impurities or grain boundaries are already present in the primary materials, the nucleation process could occur more easily close to solid-liquid interfaces. Future experiments addressing this issue –multi-step pathways of heterogeneous nucleation- should be designed to give a more complete view of this complex system.